\documentclass[10pt]{iopart}
\input amssym.def
\input amssym
\usepackage{graphicx}
\usepackage{horowitzIOP}

\begin{document}

\title{Testing AdS/CFT Drag and pQCD Heavy Quark Energy Loss}

\author{W.\ A.\ Horowitz}

\address{Frankfurt Institute for Advanced Study (FIAS),
60438 Frankfurt am Main, Germany}

\address{Department of Physics, 
Columbia University,
538 W.\ 120$^\mathrm{th}$ St.,
New York, NY 10027, USA}

\ead{horowitz@phys.columbia.edu}

\author{M.\ Gyulassy}

\address{Frankfurt Institute for Advanced Study (FIAS),
60438 Frankfurt am Main, Germany}

\address{Department of Physics, 
Columbia University,
538 W.\ 120$^\mathrm{th}$ St.,
New York, NY 10027, USA}

\ead{gyulassy@phys.columbia.edu}

\begin{abstract}
We present charm and bottom nuclear modification factors for RHIC and LHC using Standard Model perturbative QCD and recent AdS/CFT string drag energy loss models.  We find that extreme extrapolations to LHC mask potential experimentally determinable differences in the individual $R_{AA}$s but that their ratio, $R_{AA}^c/R_{AA}^b$ as a function of transverse momentum is a remarkably robust observable for finding deviations from either theoretical framework.
\end{abstract}
\vspace{-.2in}
\vspace{-.2in}
\submitto{\jpg}
\vspace{-.2in}
\section{Introduction}
\vspace{-.1in}
The shockingly large suppression of nonphotonic electrons \cite{data} observed at RHIC falsified the thus-far successful paradigm of perturbatively-calculable QCD radiative energy loss for high-\pt jets in heavy ion collisions.  This led to a spate of suggestions for novel energy loss mechanisms.  In this paper we investigate the possibility of experimentally verifying or ruling out two broad classes of models: pQCD-based derivations of radiative and collisional and AdS/CFT string drag energy loss.

While the traditional pQCD approach seemed to describe the light quark and gluon jet suppression, as measured by pions and etas with the null control of direct photons \cite{data}, data on the seemingly anomalously small viscosity to entropy ratio, intermediate-\pt \vtwocomma, and nonphotonic electron suppression suggest that, at best, it does not provide a complete understanding of heavy ion collisions \cite{Teaney:2003kp,data}.  On the other hand, the AdS/CFT paradigm, which requires a double conjecture that not only supersymmetric Yang-Mills (SYM) is dual to Type IIb string theory but that QCD is not too dissimilar to SYM, has had a number of qualitative successes: the reduction of entropy compared to the Stefan-Boltzmann (SB) limit persistent out to several times $T_c$, a small viscosity to entropy ratio ($\eta/s$), presence of Mach cones, and a large suppression of high-\pt heavy quarks \cite{strings,stringdrag}.
%
%
To further test heavy quark energy loss mechanisms and the AdS framework, we predict the individual charm and bottom $R_{AA}^Q(\eqnpt)$ for LHC and RHIC \cite{me}.  Since QCD and SYM are not identical, e.g.\ SYM has extra degrees of freedom, nonrunning coupling, etc., we are motivated to seek ratios of observables such that the differences in the two theories effectively cancel.
\vspace{-.1in}
\section{Comparison}
\vspace{-.1in}
The string drag on a heavy quark of mass $M_Q$ moving with constant velocity in a constant temperature SYM plasma in the limit of $\lambda = \sqrt{g_{SYM}^2N_c} \gg 1$, $N_c \gg 1$, $M_Q \gg T^{SYM}$ is 
$d\eqnpt/d\eqnpt = -\big(\pi\surd\lambda T^2/2M_q\big)\eqnpt$ \cite{stringdrag}.
This is quite similar to the Bethe-Heitler limit of incoherent, radiative energy loss, which goes as $d\eqnpt/dt\sim -(T^3/M_q^2) \eqnpt$, but is completely different from the usual LPM-dominated pQCD result that goes as $d\eqnpt/dt\sim -LT^3\log(\eqnpt/M_q)$.  

Using the Jacobian approximation for the nuclear modification factor, $R_{AA}(\eqnpt) = \int_\mathrm{geom} (1-\epsilon)^{n(\eqnpt)}$ where $p_{T,f} = (1-\epsilon)p_{T,i}$ and the hard production spectrum goes as $dN/d\eqnpt \sim 1/\eqnpt^{n(\eqnpt)+1}$ and the integration is over the realistic distribution of production points and medium, we may gain an intuitive understanding of our results.  For LHC $n$ is a monotonic, slowly increasing function of transverse momentum.  In this case the $\epsilon_\mathrm{pQCD}\sim\log(\eqnpt)/\eqnpt$ goes to zero faster than the increase in $n$ can compensate, and $R_{AA}^\mathrm{pQCD}$ increases with \ptcomma.  Since $\epsilon_\mathrm{AdS}$ is independent of \pt we expect $R_{AA}^\mathrm{AdS}$ to decrease with momentum as $n$ increases.  
\begin{figure}[htb!]
\vspace{-.15in}
\begin{center}
\leavevmode
\includegraphics[width=.66\columnwidth]{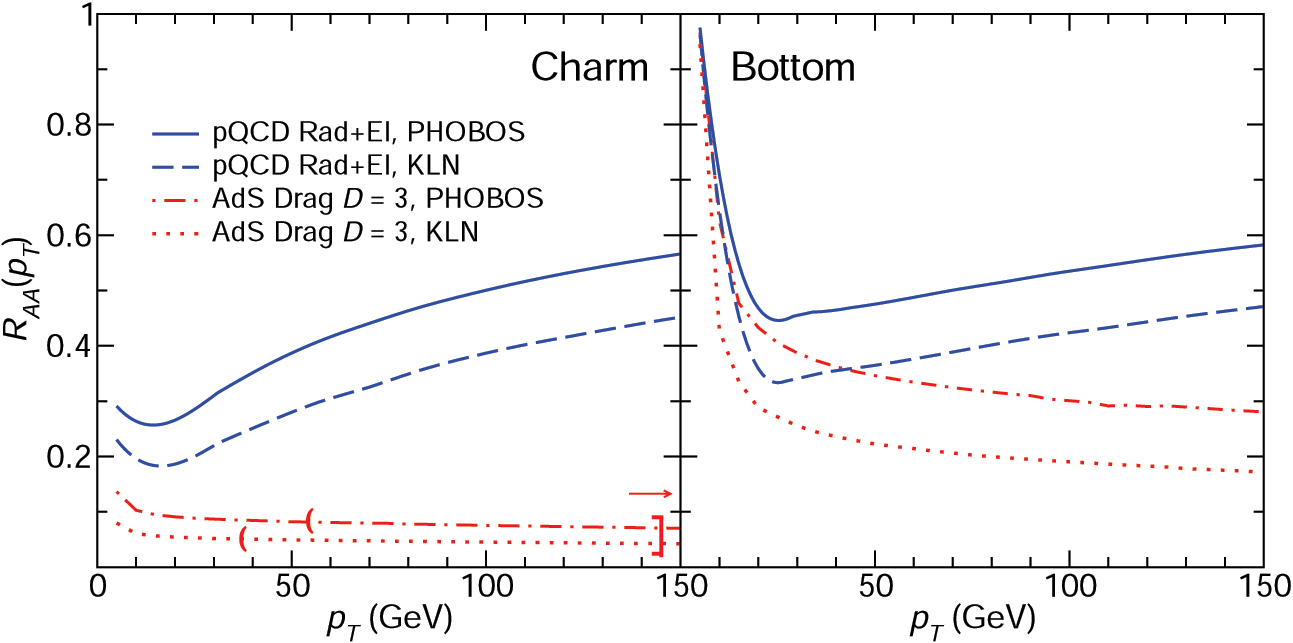}
\vspace{-.15in}
\caption{
\label{fig1}
\captionsize{$R^Q_{AA}(\eqnpt)$ for charm and bottom at LHC.  The representative pQCD curves increase with \pt while the AdS drag curves decrease.  Not shown are other samples of the range of input parameters that yield pQCD and AdS results difficult to distinguish experimentally.
}}
\end{center}
\vspace{-.2in}
\end{figure}


Comparing string drag model calculations to current and future data requires a prescription for mapping parameters in QCD to those in SYM.  We take a small sample of these possibilities with two: an ``obvious'' mapping, which takes $\alpha_s = \alpha_{SYM}$ fixed (we use a pQCD/hydro inspired value of .3 and a $D/2\pi T\sim3$ inspired value of .05) and $T_{QCD}=T_{SYM}$, and an ``alternative'' mapping with $\lambda\sim5.5$ and $T_{SYM}\approx T_{QCD}/3^{1/4}$ \cite{Gubser:2006qh}.  The ``alternative'' mapping comes from matching the $q\bar{q}$ force in AdS to that found on the lattice and equating the QCD and SYM energy densities.  We compare these predictions to the WHDG model of convolved inelastic and elastic energy loss with fixed $\eqnalphas=.3$ and to WHDG with collisional energy loss turned off, approximating the proposed large $\eqnqhat\sim 40-100$ radiative only extrapolation to LHC.  We use the PHOBOS extrapolation, $dN_g/dy = 1750$, and the KLN model of the CGC, $dN_g/dy = 2900$, as representative values for the possible range of medium densities at LHC \cite{LHCmedium}.  Our results for a sample of the $R_{AA}^Q$ curves are shown in \fig{fig1}; our naive expectations are borne out in the full numerics: the pQCD curves have a significant rise and the AdS curves fall with \ptcomma.  We note that the use of diffuse Woods-Saxon nuclear geometry and Bjorken expansion allows for nonfragile $R_{AA}^c(\hat{q}=40,100) \sim .15, .02$.  

\begin{figure}[htb!]
\vspace{-.15in}
\begin{center}
\leavevmode
\includegraphics[width=.65\columnwidth]{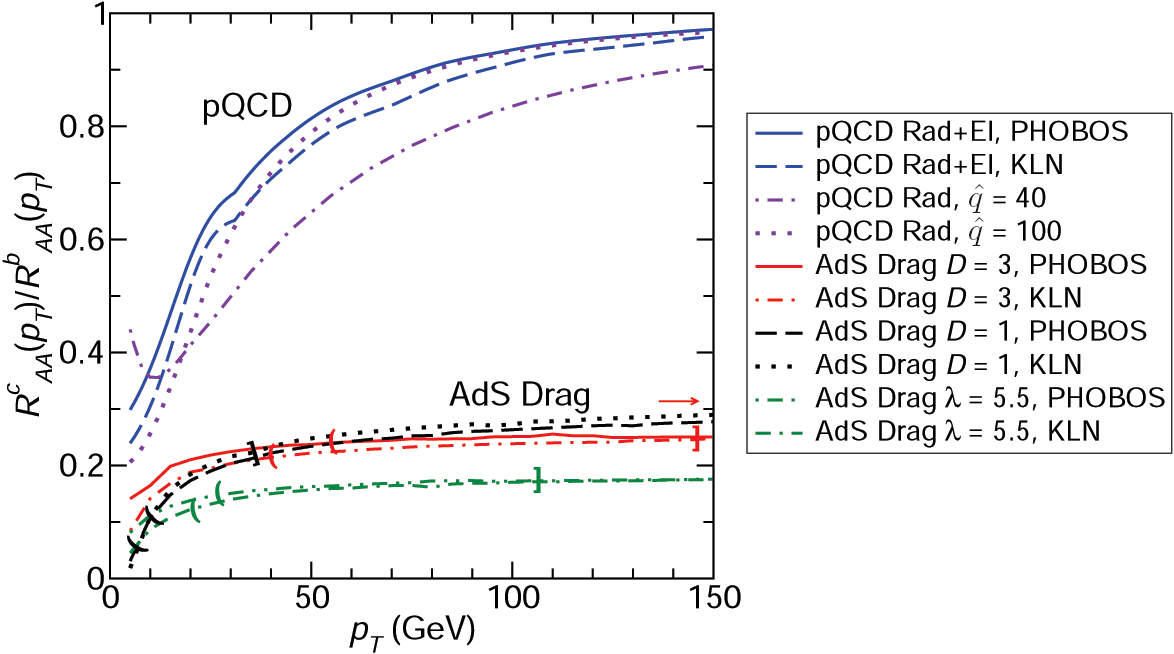}
\vspace{-.15in}
\caption{
\label{ratio}
\captionsize{Taking the ratio $R^{cb}=R^c_{AA}/R^b_{AA}$ groups the predictions into a pQCD and an AdS string drag bunch, nearly irregardless of input parameters used.}
}
\end{center}
\vspace{-.2in}
\end{figure}

Unfortunately the range of possible $R_{AA}^Q(\eqnpt)$ predictions makes experimental differentiation difficult.  We find an enhanced signal by exploiting the interplay of the quark mass and jet momentum scales by taking the ratio of 
the two, $R^{cb}(\eqnpt)=R^c_{AA}(\eqnpt)/R_{AA}^b(\eqnpt)$.  For pQCD $R^{cb}(\eqnpt) \sim 1-\eqnalphas n(\eqnpt) L^2 \log(M_b/M_c)(\eqnqhat/\eqnpt)$ asymptotically; the ratio approaches 1 from below at \highptcomma, and the approach is slower for larger quenching.  In the string drag model $R^{cb}(\eqnpt) \sim M_c/M_b$: significantly below 1, independent of momenta.  This behavior is seen in the full numerical results in \fig{ratio}.  

\begin{figure}[htb!]
\vspace{-.15in}
\begin{center}
\leavevmode
$\begin{array}{c@{\hspace{.00in}}c}
\includegraphics[width=.334\columnwidth]{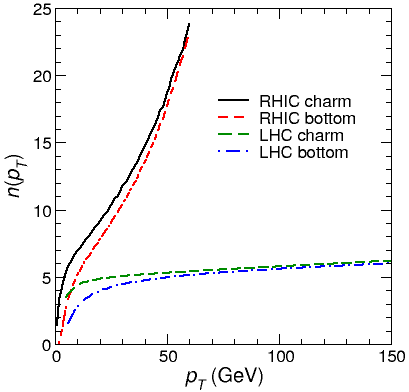} & 
\includegraphics[width=.66\columnwidth]{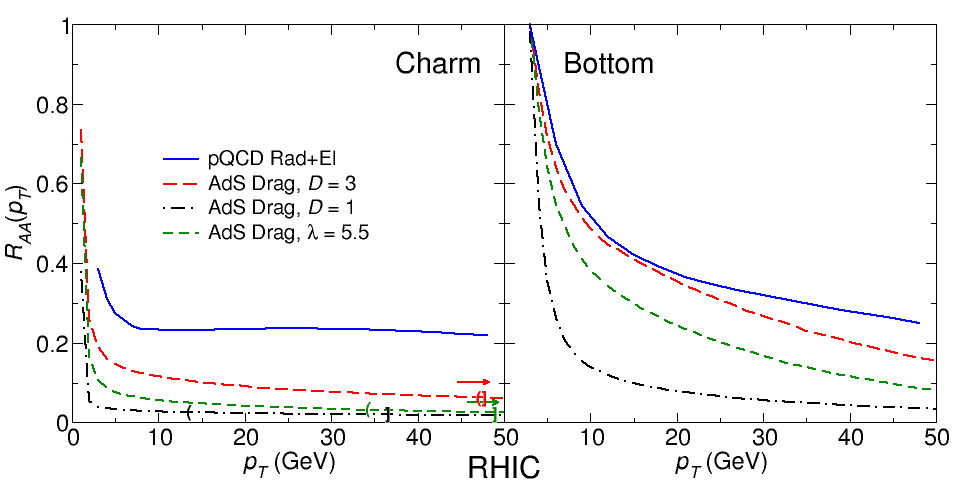} \\ [-0.in]
{\mbox {\scriptsize {\bf (a)}}} & {\mbox {\scriptsize {\bf (b)}}}
\end{array}$
\vspace{-.15in}
\caption{
\label{rhic}\captionsize{(a) The power law production index $n(\eqnpt)$ for RHIC and LHC, relatively momentum independent for LHC but harder and \ptcomma-dependent at RHIC. (b) \raacpt and \raabpt for RHIC for WHDG pQCD and AdS string drag.  The large increase in $n(\eqnpt)$ overcomes the decreasing fractional momentum loss for pQCD; both \ads drag and pQCD results decrease as a function of momentum at RHIC.}
}
\end{center}
\vspace{-.2in}
\end{figure}


One expects nonperturbative corrections to pQCD calculations as momenta decreases; the finite scale at which these become large is numerically uncertain at experimentally accessible energies.  Contrariwise the AdS/CFT string drag formula appears to require corrections for large $\gamma$.  Requiring a time-like string endpoint for finite mass heavy quarks gives a ``speed limit'' of $\gamma_c = (1+2M_q/\surd\lambda T)^2 \sim 4M_q^2/\lambda T^2$.  Note that this is again parametrically identical to the limiting momentum for going from Bethe-Heitler to LPM loss perturbatively with one less power of $T/M_q$.  To get a sense of these scales we include a ``('' for the smallest possible $\gamma_c$, $\gamma_c\big(T(\vec{x}_\perp=\vec{0},\tau=\tau_0)\big)$, and ``]'' for the largest, $\gamma_c\big(T_c\big)$ in \fig{ratio}.

\begin{figure}[htb!]
\vspace{-.15in}
\begin{center}
\leavevmode
\includegraphics[width=.66\columnwidth]{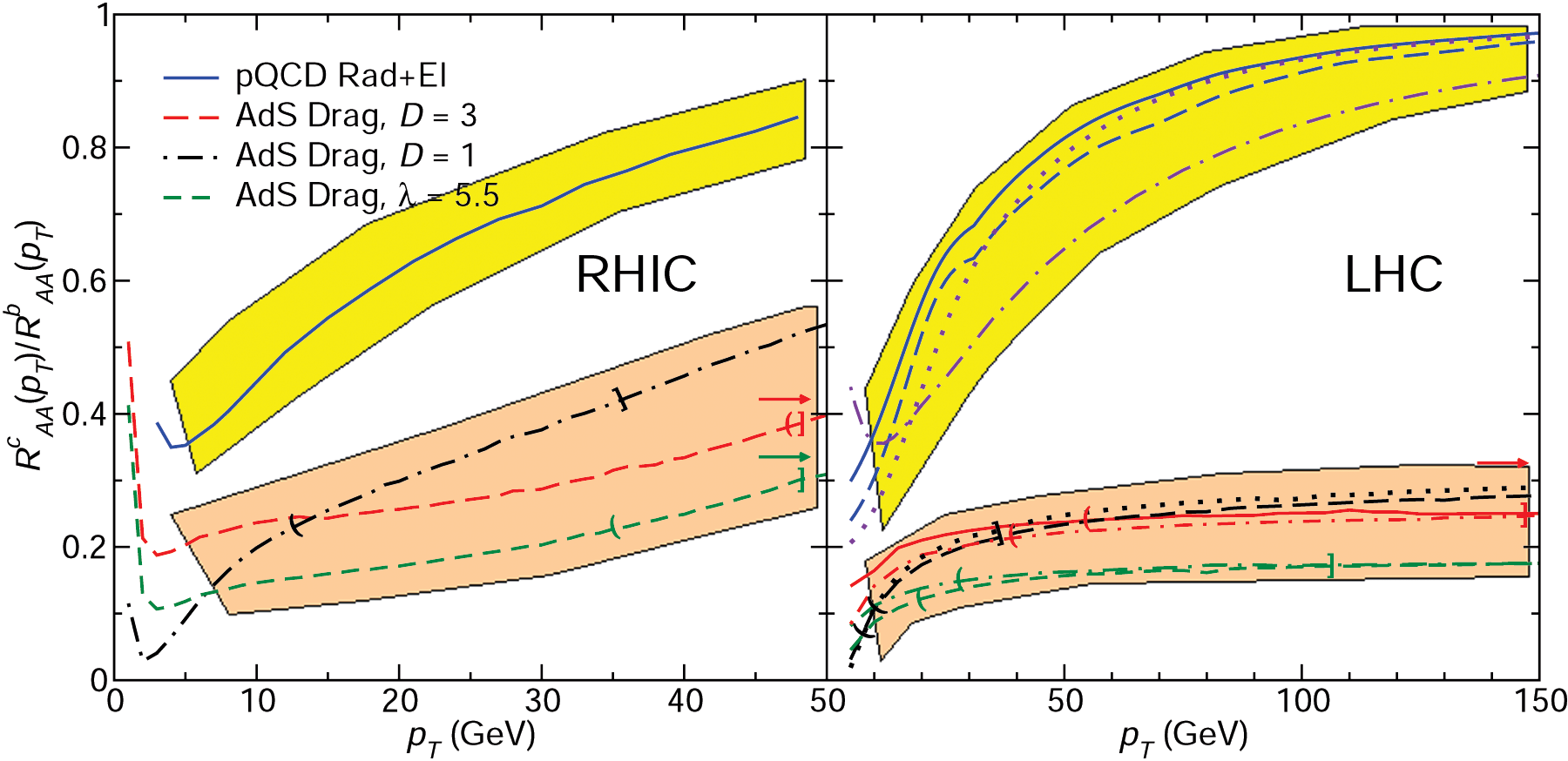}
\vspace{-.15in}
\caption{
\label{rhicratio}
\captionsize{$R^{cb}$ predictions for RHIC using WHDG pQCD and AdS string drag for a range of input parameters.  While the hardening of the production spectrum reduces the dramatic bunching at RHIC as compared to LHC, the lower temperature at RHIC means the \ads drag formalism is applicable up to higher momenta.  Note that $R^{cb}$ is plotted only to 50 GeV for RHIC.}
}
\end{center}
\vspace{-.2in}
\end{figure}

Future detector upgrades at RHIC should allow individual measurements of $c$ and $b$ quarks.  We show predictions for the individual $R^Q_{AA}(\eqnpt)$ in \fig{rhic} (b) and for the double ratio in \fig{rhicratio}.  Since the production spectra are much harder and have significant \pt dependence, \fig{rhic} (a), one does not see scaling as cleanly as in the LHC predictions.  An advantage of RHIC, however, will be its lower multiplicity and hence medium temperature: corrections to the string drag energy loss will occur at higher momenta.
\vspace{-.1in}
\section{Conclusions}
\vspace{-.1in}
RHIC and LHC predictions of $R^Q_{AA}$ for charm and bottom quarks using pQCD and AdS string drag energy loss were found.  
Reasonable input parameters lead to different \pt dependencies for the two model classes, but extreme extrapolations to LHC mask the results.  Examining the ratio $R^{cb}=R^c_{AA}/R^b_{AA}$ cancels much of the dependence on input parameters and groups the predictions into pQCD and AdS bunches, especially at LHC.  However a thorough understanding of the regions of theoretical self-consistency will be crucial for strong experimental statements to be made.

\end{document}